\documentclass[%
 amsmath,amssymb,
 aps,
twocolumn
]{revtex4-2}

\usepackage [english]{babel}
\usepackage [autostyle, english = american]{csquotes}
\MakeOuterQuote{"}

\usepackage{appendix}

\usepackage{graphicx}
\usepackage{dcolumn}
\usepackage{bm}
\usepackage{hyperref}
\usepackage[mathlines]{lineno}

\usepackage[
margin=1in,
]{geometry}

\usepackage[dvipsnames]{xcolor}

\hypersetup{
    colorlinks,
    linkcolor={red!50!black},
    citecolor={blue!50!black},
    urlcolor={blue!80!black}
}

\usepackage{soul}
\usepackage[normalem]{ulem} 
\usepackage{chemmacros}

\makeatletter
\newcommand*{\rightharpoonupfill@}{%
  \arrowfill@\relbar\relbar\rightharpoonup
}

\newcommand*{\leftharpoondownfill@}{%
  \arrowfill@\leftharpoondown\relbar\relbar
}

\newcommand{\xrightleftharpoons}[2][]{%
  \ensuremath{%
    \mathrel{%
      \settoheight{\dimen@}{\raise 2pt\hbox{$\rightharpoonup$}}%
      \setlength{\dimen@}{-\dimen@}%
      \edef\CA@temp{\the\dimen@}%
      \settoheight\dimen@{$\rightleftharpoons$}%
      \addtolength{\dimen@}{\CA@temp}%
      \raisebox{\dimen@}{%
        \rlap{%
          \raisebox{2pt}{%
            $%
            \ext@arrow 0359\rightharpoonupfill@{\hphantom{#1}}{#2}%
            $%
          }%
        }%
        \hbox{%
          $%
          \ext@arrow 3095\leftharpoondownfill@{#1}{\hphantom{#2}}%
          $%
        }%
      }%
    }%
  }%
}
\makeatother

\DeclareMathOperator{\Tr}{Tr}

\usepackage[dvipsnames]{xcolor}
\usepackage{soul}

\begin{document}

\title{Numerical Gate Synthesis for Quantum Heuristics\\ on Bosonic Quantum Processors}

\author{A. Bar\i\c{s} \"Ozg\"uler}
\affiliation{%
  Fermi National Accelerator Laboratory, Batavia, Illinois, 60510
}%
\affiliation{%
  Superconducting Quantum Materials and Systems Center (SQMS), Fermilab
}%
\author{Davide Venturelli}
\affiliation{%
  Quantum Artificial Intelligence Laboratory (QuAIL), NASA Ames Research Center
}%
\affiliation{%
  USRA Research Institute for Advanced Computer Science (RIACS)
}%
\affiliation{%
  Superconducting Quantum Materials and Systems Center (SQMS), Fermilab
}%

\date{August 7, 2022}

\begin{abstract}

There is a recent surge of interest and insights regarding the interplay of quantum optimal control and variational quantum algorithms. We study the framework in the context of qudits which are, for instance, definable as controllable electromagnetic modes of a superconducting cavity system coupled to a transmon. By employing recent quantum optimal control approaches described in (Petersson and Garcia, 2021), we showcase control of single-qudit operations up to eight states, and two-qutrit operations, mapped respectively onto a single mode and two modes of the resonator. We discuss the results of numerical pulse engineering on the closed system for parametrized gates useful to implement Quantum Approximate Optimization Algorithm (QAOA) for qudits. The results show that high fidelity ($>$ 0.99) is achievable with sufficient computational effort for most cases under study, and extensions to multiple modes and open, noisy systems are possible. The tailored pulses can be stored and used as calibrated primitives for a future compiler in circuit quantum electrodynamics (cQED) systems. 

\end{abstract}

\maketitle

\section{\label{sec:Introduction}Introduction}

Quantum information processing arbitrates controlled interaction of the Hilbert space of a quantum system, for the purpose of generating a target probability distribution expressed in a computational basis defined by an experimental measurement scheme. The Hilbert space of a quantum system generally grows exponentially with the number of degrees of freedom, but for the purpose of quantum information processing it needs to be opportunistically partitioned in order to execute algorithms. The most common encoding exploits a collection of qubits, two-level systems, and it is known that the dimensionality of the Hilbert space is maximized when the states are arranged as a collection of qutrits, three-level systems, for a fixed number of allowed quantum states 
 \cite{greentree2004maximizing}. 
Without loss of generality, multiple qudits can be merged into the definition of a new qudit, and a qudit can be mapped via binary encodings into a minimum of $\log_2(d)$ qubits and viceversa. For instance, the binary expansion

\begin{eqnarray}
\bigotimes_{k=0}^{N-1}|s_k\rangle& 
\xrightleftharpoons[qubits]{qudits}
&\left|\sum_{k=0}^{N-1} s_k d^k\right\rangle
\label{eq:quditmapping}
\end{eqnarray}
would map the computational basis state of a ququart (i.e. a four-level qudit) onto two qubits $|0\rangle\rightarrow|00\rangle$, $|1\rangle\rightarrow|01\rangle$, $|2\rangle\rightarrow|10\rangle$, $|3\rangle\rightarrow|11\rangle$.

It is known since the beginning of quantum computing architecture research that universal quantum computing could be achieved by operating constructively on single-qubit and two-qubit at a time~\cite{divincenzo1995two} via the implementation of quantum gates temporally arranged into quantum circuits. A similar result is known for qudits of arbitrary dimension~\cite{bullock2005asymptotically,wang2020qudits}, which can provide hardware-efficient solutions \cite{liu2021constructing} and lower-depth gate compilation and noise improvement compared to qubit-based systems \cite{gokhale2019asymptotic, otten2021impacts, blok2021quantum, gustafson2021prospects, gustafson2022noise}. Of particular interest in the current period of technological maturity of quantum processors (the NISQ Era~\cite{preskill2018quantum}) are variational algorithms such as the Quantum Approximate Optimization Algorithm (QAOA) and the Variational Quantum Eigensolver (VQE) that might achieve some quantum advantage without the fault-tolerance overhead of active error-correction~\cite{cerezo2021variational}. Typically the quantum circuits of these algorithms feature unitary gates implementing a set of parametrized single-qudit rotations $U_M(\beta)$ depending on some real angle $\beta$. For instance, let us consider the set of $SU(2)$ rotations around the $X$-axis of the Bloch sphere for qubit systems, and the set of $SO(3)$ rotations that leave invariant the $|0\rangle+|1\rangle+|2\rangle$ state for qutrits. Their matrix representations  $U_{M}^{(2)}(\beta)$ and $U_{M}^{(3)}(\beta)$ are, respectively:
\setlength{\thickmuskip}{0mu}
\setlength{\medmuskip}{0mu}
\begin{eqnarray}
    U_{M}^{(2)}&\equiv&\begin{bmatrix} c_{\frac{\beta}{2}}&-is_{\frac{\beta}{2}} \\is_{\frac{\beta}{2}}&c_{\frac{\beta}{2}}\end{bmatrix}\label{eq:UMix}\\
    U_{M}^{(3)}&\equiv&\frac{1}{3}\begin{bmatrix}
     1\text{+}2c_\beta & 1-c_\beta-\sqrt{3}s_\beta & 1-c_\beta\text{+}\sqrt{3}s_\beta\\
     1-c_\beta\text{+}\sqrt{3}s_\beta & 1\text{+}2c_\beta & 1-c_\beta-\sqrt{3}c_\beta\\
     1-c_\beta-\sqrt{3}s_\beta & 1-c_\beta\text{+}\sqrt{3}s_\beta & 1\text{+}2c_\beta,\nonumber
    \end{bmatrix}
\end{eqnarray}
\setlength{\thickmuskip}{2mu}
\setlength{\medmuskip}{2mu}
where $c_x$, $s_x$ indicate $\cos(x)$ and $\sin(x)$ and the computational basis states are ordered in the canonical ascending way.
The two-qudit gates of interest for QAOA/VQE ans\"atze are often diagonal in the computational basis. For instance, the following two-qudit and two-qutrit unitary gates $U_C(\gamma)$ introduce a phase shift by the angle $\gamma$ if the two qudits have the same computational state:
\begin{eqnarray}
    U_C^{(2)}&\equiv&\begin{bmatrix}
    e^{i\gamma} & 0 & 0 & 0\\
    0 & 1 & 0 & 0\\
    0 & 0 & 1 & 0\\
    0 & 0 & 0 & e^{i\gamma}
    \end{bmatrix}\label{eq:UCost}\\
    U_C^{(3)}&\equiv&
    \begin{bmatrix}
    e^{i\gamma} & 0 & 0 & 0 & 0 & 0 & 0 & 0 & 0\\
    0 & 1 & 0 & 0 & 0 & 0 & 0 & 0 & 0\\
    0 & 0 & 1 & 0 & 0 & 0 & 0 & 0 & 0\\
    0 & 0 & 0 & 1 & 0 & 0 & 0 & 0 & 0\\
    0 & 0 & 0 & 0 & e^{i\gamma} & 0 & 0 & 0 & 0\\
    0 & 0 & 0 & 0 & 0 & 1 & 0 & 0 & 0\\
    0 & 0 & 0 & 0 & 0 & 0 & 1 & 0 & 0\\
    0 & 0 & 0 & 0 & 0 & 0 & 0 & 1 & 0\\
    0 & 0 & 0 & 0 & 0 & 0 & 0 & 0 & e^{i\gamma}\\
    \end{bmatrix},\nonumber
\end{eqnarray}
where the canonically ordered basis for the matrix representation is used~\footnote{$|00\rangle$, $|01\rangle$, $|10\rangle$, $|11\rangle$ for qubits, and $|00\rangle$, $|01\rangle$, $|02\rangle$, $|10\rangle$, $|11\rangle$, $|12\rangle$, $|20\rangle$, $|21\rangle$, $|22\rangle$ for qutrits}. Note that for the most common case of qubits $U_C^{(2)}\propto\exp(i (\gamma/2)\sigma_z\otimes\sigma_z)$, where $\sigma_z$ are the standard Pauli matrices. For circuit quantum electrodynamics (cQED) systems, note also that there are  ways to find effective spin models, which is generally used for encoding of quantum heuristic algorithms \cite{miyazaki2022effective}.

Implementing parametrized gates such as (\ref{eq:UMix}-\ref{eq:UCost}) starting from the elementary interactions provided by a NISQ  processor is a non-trivial problem of \emph{synthesis}~\cite{magann2021pulses}, which often can be tackled only via heuristic numerical approaches and online experimental calibration~\cite{klimov2020snake}. In this work, we consider the problem of synthesis of gates of the type (\ref{eq:UMix}-\ref{eq:UCost}) by driving with carefully optimized time-dependent interactions in a system of interacting states.
More specifically, the Hilbert space we are considering is spanned by a truncated set of anharmonic bosonic modes, defined with second quantized operators $a_m$, coupled in a density-density fashion. The corresponding many-body Hamiltonians and their truncated diagonal first quantization representations are:
\begin{eqnarray}
H_m      &=& \omega_m a_m^\dagger a_m + \xi_m (a_m^\dagger a_m)^2\label{eq:Hm}\\
         \Big|_{n_m\ }&\xrightarrow{}&|0\rangle \langle 0|+\sum_{n=1}^{n_m-1} \left[\omega_m n + \xi_m n^2\right] |n\rangle \langle n|,\nonumber\\
H_{mm\prime}^{int} &=&  \xi_{mm^\prime} a_m^\dagger a_m  a^\dagger_{m^\prime} a_{m\prime}\label{eq:Hmm}\\
             \Big|_{{n_m}\atop{n_{m^\prime}}}&\xrightarrow{}&|00\rangle \langle 00|+\sum_{n=1}^{n_m-1} \sum_{k=1}^{n_{m\prime}-1} \xi_{mm^\prime}nk |nk\rangle\langle nk|,\nonumber
\end{eqnarray}
where $n_m$, $n_{m\prime}$ are the number of levels considered for each mode. In photonic implementations, $\xi_m$ is called the self-Kerr coefficient for mode $m$ and $\xi_{mm^\prime}$ is called the cross-Kerr coefficient between modes $m$ and $m^\prime$. 
\begin{figure}[!htbp]
\begin{centering}
\includegraphics[width=\columnwidth]{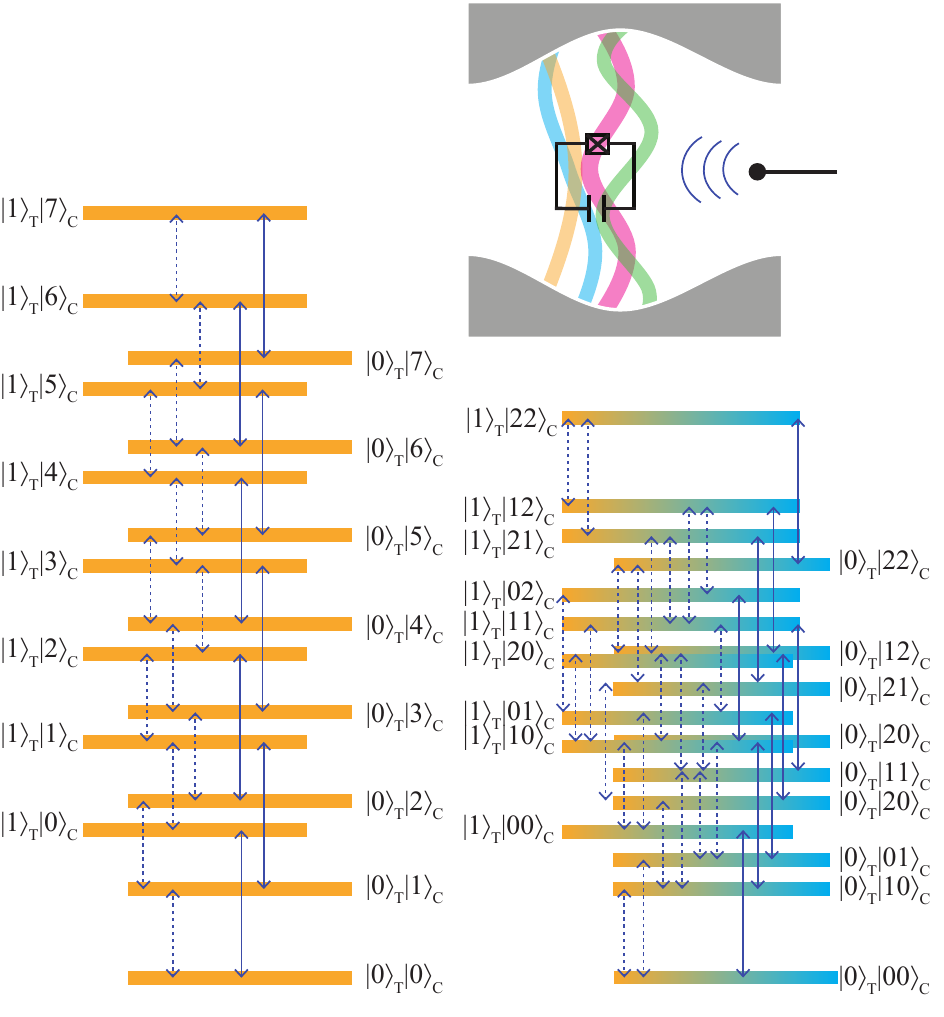}
\par\end{centering}
\caption{\label{fig:System_and_Spectrum}
Top: System figure with colored waves representing different cavity electromagnetic modes. Eigenspectrum for $\mathcal{H}^{(A)}$ (left) and $\mathcal{H}^{(B)}$ (right). 
System parameters and resonant frequencies are given in Section \ref{sec:quantum_control}.
Arrows indicate the transition frequencies.
Dashed (continuous) arrows represent transitions between different energy levels with $|0\rangle_{T}$ and $|0\rangle_{T}$ ($|1\rangle_{T}$).}
\end{figure}
We are considering two illustrative setups in order to describe how quantum information could be manipulated in systems featuring Hamiltonians of the type (\ref{eq:Hm}-\ref{eq:Hmm}). 
In particular we consider that there is one "control qubit mode" $T$ whose Hilbert space is truncated to the first two computational states: 
\begin{eqnarray}
H_T &=& |0\rangle_T \langle 0|_T+(\omega_T+\xi_T)|1\rangle_T \langle 1|_T,\label{eq:HT}
\end{eqnarray}
and either one or two computational modes ($C$) interacting with the control mode, respectively truncated to the first 8 and 3 computational states:
\begin{eqnarray}
\mathcal{H}^{(A)} &=& H_T + H_m + H_{Tm}^{int}\Big|_{\substack{n_T=2\\n_m=8}}
\label{eq:H2}
\\
\mathcal{H}^{(B)} &=& \left.  \begin{array}{l}   
                      H_T + H_l + H_m\\
                      + H_{Tl}^{int} + H_{Tm}^{int}                      \nonumber\\                      \end{array}\right|_{^{\substack{n_T=2\\n_l=3\\n_m=3}}},
\end{eqnarray}
where the dependence over the $\omega$ and $\xi$ parameters of the Hamiltonians is implied. This setup is a specific case of a generalized Jaynes-Cummings model~\cite{blais2021circuit}. Note that for $\mathcal{H}^{(B)}$, each $C$ mode is naturally a qutrit, while as noted in Eq.~(\ref{eq:quditmapping}), the quantum occupation numbers of the cavity modes could be directly associated to qubit registers via binary expansion~\cite{sawaya2020resource}. 

In Fig.~\ref{fig:System_and_Spectrum}, we show the energy spectrum of two specifications of $\mathcal{H}^{(A)}$ and $\mathcal{H}^{(B)}$ as well as a pictorial representation of a possible experimental setup that could be described by such effective Hamiltonians: A transmon circuit is embedded into a multimode 3D superconducting cavity, driven by the field of a coupled antenna. Indeed, our reference Hamiltonians can be derived by considering the superconducting transmon to be coupled with the cavity resonator in a dispersive way, i.e., by considering the effective interaction derived by perturbation theory assuming that the ratio of the transmon-cavity coupling and the difference between the transmon and the cavity fundamental frequencies is small, and neglecting the small effective couplings (i.e. cross-Kerr) between the cavity modes~\cite{blais2021circuit, ma2021quantum}. The quantum control drive can be introduced in the model by adding a time-dependent term that allows to create and destroy excitations of a mode $m$:
\begin{equation}
    H^{drive}_{m}(t) = d_m(t) a_m + \bar{d}_m(t) a_m^\dagger,\label{eq:HD}
\end{equation}
where $d_m(t)$ are complex functions. This control Hamiltonian could be related to the (comparatively slowly varying) field generated by the antenna via phenomenologically justified approximations~\cite{gerry2005introductory}. 


Having introduced the main definitions and the systems under study, we outline the rest of the paper. In section \ref{sec:quantum_control} we present the synthesis problem from a numerical point of view, following the implementation of quantum optimal control numerics in the open source package \texttt{Juqbox.jl}~\cite{Juqbox_Github}. Subsection \ref{subseq:evaluation} will present results for the synthesis of simple QAOA proof-of-concept circuits based on the parallel execution of gates (\ref{eq:UMix})-(\ref{eq:UCost}). Finally,
in Section \ref{sec:discussion} 
we will discuss future work, including improvements and generalizations of our case study to larger and realistic systems and what is needed for this method to be applied in practice for compilation of variational quantum algorithms in realistic bosonic quantum processors based on 3D cQED technology.

\section{\label{sec:quantum_control} Pulse Engineering Approach}

The \emph{gate synthesis} problem that we are facing could be framed as the task of discovering the functions $d_m(t)$ that allow the Schr\"odinger evolution for a time $\tau$ of $\mathcal{H}+H^{drive}$ to match as close as possible a target unitary operation $U$: 
\begin{eqnarray}
    U &\;\simeq\;& \mathcal{U}(\tau)=\mathcal{T} \exp\left[-\frac{i}{\hbar}\int_0^\tau dt\left( \mathcal{H}+H^{drive}(t)\right)\right]
    \label{eq:schroevol}
\end{eqnarray}
In particular, as discussed in the previous section, we will be considering Eqs.(\ref{eq:H2}) for $\mathcal{H}$ and Eqs. (\ref{eq:UMix}-\ref{eq:UCost}) as target unitary matrices. In order to solve synthesis numerically, the problem is cast into an  optimization challenge over a finite number of real parameters, which can be tackled following the theory of quantum optimal control (QOC)~\cite{palao2002quantum}. There are multiple strategies currently implemented for gate sythesis via QOC or machine learning, all with respective benefits and tradeoffs. However, these methods are currently tested on specific limited cases, and insights are difficult to generalize, e.g. see~\cite{riaz2019optimal, niu2019universal, PRXQCTRL}. In this paper we follow the techniques described in Ref.~\cite{petersson2021optimal}, targeting specifically cQED models, which we will now briefly review and contextualize for the system under study. 

We leverage a key simplification of the QOC problem, consisting in the decomposition of the $d_m(t)$ control functions into a truncated basis spanned by a linear combination of $N_b$ B-spline quadratic polynomials, $S_b(t)$, corresponding to wavelets modulated with $N_f$ resonant frequencies, i.e.
\begin{eqnarray}
    d_m(t)&=&\sum_k^{N_f} e^{i\Omega_{m,k}} W_{m,k}(t)\nonumber\\
    W_{m,k}(t)&=&\sum_b^{N_b}\alpha_{m,k,b}S_b(t),
    \label{eq:Bsplines}
\end{eqnarray}
where $\alpha$s are complex coefficients, representing the unknowns of the optimization problem. The choice of B-splines as a basis for expansion is motivated by computational efficiency of parametrization of the control functions.
The resonant frequencies $\Omega_{m,k}$ are defined by considering the energy differences between the states corresponding to the creation or annihilation of a boson, leaving the remaining occupations unchanged. Signals tuned at these frequencies initiate transitions as it can be proven by first order time-dependent perturbation theory.

We show in Fig.~\ref{fig:System_and_Spectrum} the resonant frequencies for our illustrative systems: for $\mathcal{H}^{(A)}$, we count 8 transitions related to $T$-bosons and 14 transitions for $C$-bosons for a total of 22 resonant frequencies. For $\mathcal{H}^{(B)}$, there are 9 resonant frequencies in total that trigger T transitions, and 24 transitions related to the C modes. However, some transitions are degenerate -- only 17 different frequencies are required. 

We consider the following values of parameters, with reference to a perspective reference cQED potential implementation: $\omega_{\mathrm{T}}/2\pi=$\,5\,GHz; $\omega_{\mathrm{m}}/2\pi=$\,3\,GHz, $\omega_{\mathrm{l}}/2\pi=$\,4\,GHz; $\xi_m/2\pi=$\,0.6\,MHz, $\xi_l/2\pi=$\,0.9\,MHz; $\xi_T/2\pi =$\,200\,MHz. In line with our inspiration of a cavity-transmon systems in the dispersive regime~\cite{nigg2012black}, we assign interaction parameters to be the geometric means of the local self-interactions $\xi_{Tm}/2\pi =  \sqrt{\xi_m \times \xi_T}/2\pi =$\,10.95\,MHz and $\xi_{Tl}/2\pi = \sqrt{\xi_l \times \xi_T}/2\pi =$\,13.42\,MHz. The parameters that we used for $\mathcal{H}^{(A)}$ and $\mathcal{H}^{(B)}$ are inspired from expectations of results that would be obtained applying black-box quantization to Tesla-cavity systems~\cite{romanenko2020three} coupled dispersively to transmons with coherence times $\simeq$ 100 $\mu s$ \cite{nersisyan2019manufacturing}. Following that inspiration, we assume small linewidth for the cavity mode compared to their separations and we set the minimum frequency difference between the transmon and cavity mode frequencies to be of the order of the GHz, in order to justify independent access of the control pulses to the transmon and for each cavity modes.


Following \texttt{Juqbox.jl}~\cite{Juqbox_Github}, the \emph{pulse engineering} algorithm attempts to discover the best $\alpha_{m,k,b}$ coefficients (i.e. $2 \times N_f \times N_b$ real parameters), which works as follows. Initially, a random pulse is selected by initializing the vector of parameters using random positive numbers uniformly distributed within ${[0, 0.2\,\text{MHz})}$. Then, an objective function is calculated (see Subsection \ref{subseq:evaluation}) and the pulse is iteratively updated by computing the Schr\"odinger evolution and gradients efficiently by symplectic time-integration of  adjoint equations~\cite{petersson2020discrete}. Note that due to the B-spline parametrization, the number of control parameters does not depend directly on the pulse total duration $\tau$. However, the number of B-splines $N_b$ defines the design of the temporal structure of the pulses, so one needs to choose large enough $\tau$ and $N_b$ to allow the method to converge to a numerically robust solution. In particular, the slowest frequency resolution of the pulses is given by 1/$\tau$. We choose to vary $\tau$ in the 500-8000 ns range for our numerical experiments on $\mathcal{H}^{(A)}$ and $\mathcal{H}^{(B)}$, allowing for a frequency resolution of 0.125-2 MHz. 
The B-splines vary on the time scale $\tau/N_b$. Hence we choose $N_b$ = 10 to allow resolution at the scale of $\xi_{Tm}$, which controls multiple energy separations in the spectrum. The values of $\xi_l$, $\xi_m$ define the smallest resonant frequencies. 

\begin{figure*}[!htbp]
\begin{centering}
\includegraphics[width = .99 \textwidth]{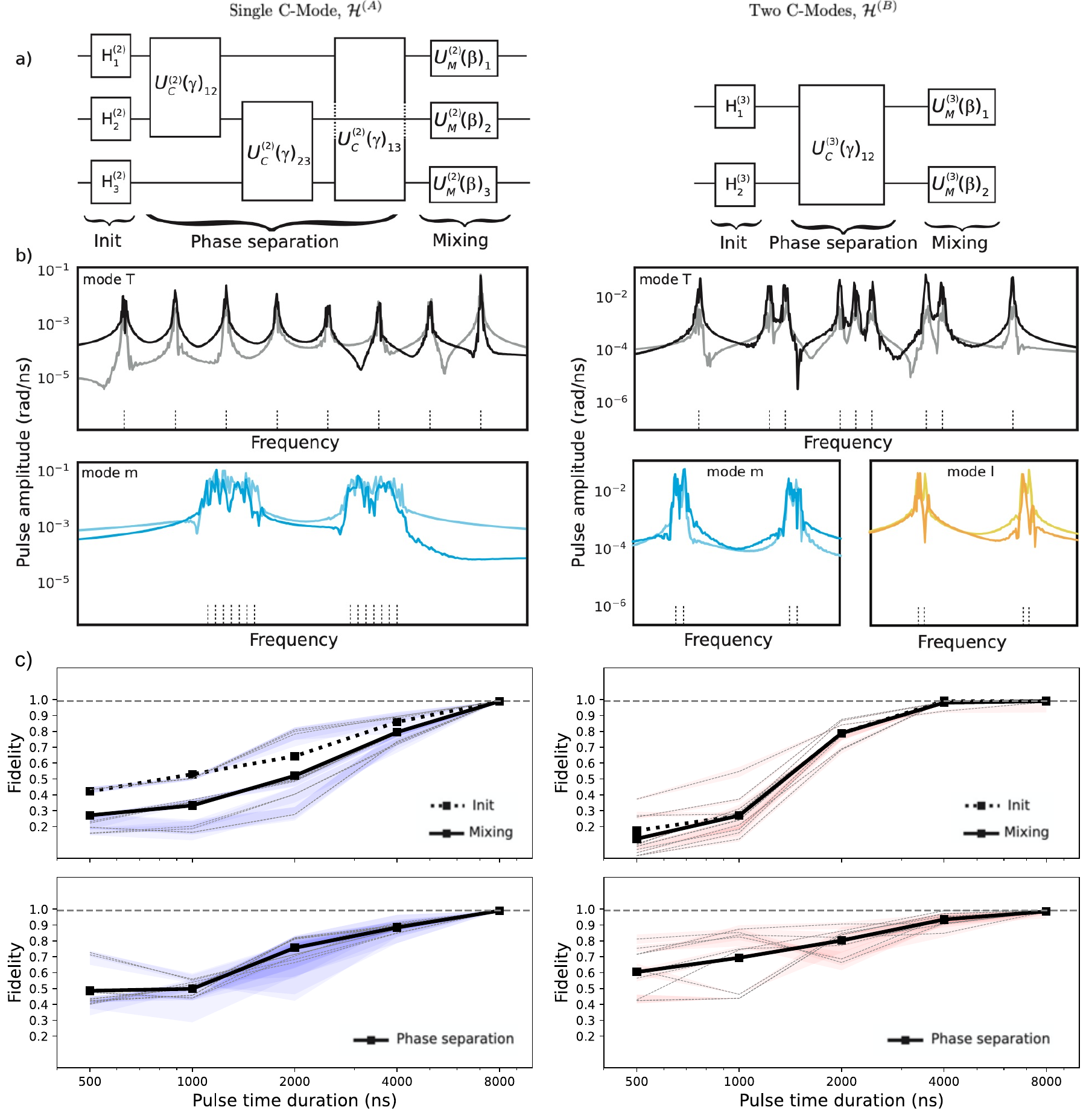}
\par\end{centering}
\caption{\label{fig:EngineeredPulses} 
(a) Prototype circuits for the synthesis of Max-k-Cut QAOA. Single C-mode represent 8 computational states (equivalent to 3 qubits).  (b) Illustrative Fourier spectrum of a high-fidelity engineered pulse via \texttt{Juqbox.jl}. Top row shows results for $d_T(t)$ while the bottom row shows the control of the computational modes ($d_m(t)$ and $d_l(t)$). Darker tones (black, blue, orange) indicate the pulses that synthesize mixing layers, while light tones (gray, cyan, yellow) refer to phase-separation layers. (c) Fidelity for pulse engineered QAOA layers of the prototype circuits. Black lines indicate the mean across angles, individually plotted in gray. Each line is the mean of 10 random restarts (20-80 percentiles across restarts is plotted as shaded area). Leakage plots are presented in the Supplementary Material.}
\end{figure*}

\subsection{Evaluation test case: QAOA}\label{subseq:evaluation}

Our numerical prototype experiment is based on the synthesis of QAOA-like quantum circuits, which in their basic implementation consist of the layered alternated application of \emph{phase-separation} unitary gates and \emph{mixing} gates~\cite{hadfield2019quantum}. With reference to the known Max-k-Cut qudit mapping of QAOA~\cite{fuchs2021efficient}, where k corresponds to the dimensionality of the qudits, we can craft the phase-separation layers using $U_C(\gamma)_{ij}$ gates and we have the freedom of designing the mixing layers using the $U_M(\beta)_{i}$ gates in Eqs.~\ref{eq:UMix}-\ref{eq:UCost}, where $i$, $j$ indicate the distinguishable qudits that are targeted by the specific gate execution. Other choices would also be appropriate~\cite{deller2022quantum}.
For clarity, in Fig.~\ref{fig:EngineeredPulses}-a, we show the two toy-model circuits that we are going to synthesize, respectively via pulse engineering on $\mathcal{H}^{(A)}$ and $\mathcal{H}^{(B)}$. For completeness, the test circuit include the \emph{initialization} operation, which is usually taken to be a generalized Hadamard gate (although it could be substituted by a mixing over the $|0\rangle^{\otimes N}$ state).

We note that quantum processor programmers have formally the freedom to execute gates sequentially or in parallel, and to exchange them in temporal execution order if they commute. However, in a real world implementation, if the processor is not fault-tolerant, under reasonable assumptions we expect decoherence and dephasing errors to be roughly proportional to execution time, so a compiler for NISQ algorithms often tries to parallelize gate execution as much as possible~\cite{venturelli2018compiling}. Moreover, considering the mapping of the computational variables to the spectrum of the Hamiltonians (Fig.~\ref{fig:System_and_Spectrum}), the possible qudit identity assignments are inequivalent with respect to pulse engineering, although it would be inconsequential if the synthesis was perfect. \texttt{SWAP} operations could restrict the number of active qudits, by relegating some states to be just memory storage and not participate in processing. However, these operations and controls for our Hamiltonians need to be synthesized as well, increasing the complexity of the entire compilation significantly. 
Bearing in mind these considerations, in our case study we choose to implement the single-qudit gates in parallel when possible, without implementing \texttt{SWAP}s but directly synthesizing all required two-body interactions instead across the entire Hilbert space. We will discuss in Section~\ref{sec:discussion} the scalability issues associated to this approach.

Noting that in cQED implementations, the Hamiltonians in Eqs.~(\ref{eq:H2}) are defined on truncated versions of a physically infinite Hilbert space, it is customary to include a few additional \emph{guard states} corresponding to high occupation of boson modes to help the robustness of the numerical optimization, i.e., the following parameters are renormalized $n_T\rightarrow\tilde{n}_T= n_T+\delta n_T$, $n_m\rightarrow\tilde{n}_m=n_m+\delta n_m$ and $n_l\rightarrow\tilde{n}_l=n_l+\delta n_l$, where $\delta n$ represent guard states with values in Table \ref{tab:parameters}.

Following \cite{petersson2021optimal}, the optimization objective to be minimized is chosen to be a sum of the infidelity and average leakage. The infidelity is a measure of a similarity score between the synthesized unitary matrix and the target, which can be defined as $O_F=1-|\Tr(\mathcal{U}(\tau)^\dagger U)/E|^2$, where E is a normalization constant. The average leakage is defined as $O_L=(1/\tau)\int_0^\tau \Tr(\mathcal{U}^\dagger(t) W \mathcal{U}(t))dt$, where $W$ is a diagonal matrix which is non-zero only on the indices corresponding to the guard levels. 
The weights in $W$ are set to be 1.0 for the highest guard state and then decrease exponentially in powers of 10 for each lower state. The objective of the numerics is to minimize $O=O_F+O_L$ by solving the related optimization problem on the $\alpha$ parameters by using the IPOPT L-BFGS optimizer~\cite{wachter2006implementation} and using the efficient \texttt{Juqbox.jl} numerical integration scheme to compute the required $O$ and $\nabla_\alpha O$. 

\begin{table}[!htbp]
\centering
\caption{\label{tab:parameters}Parameters used for prototype (See Fig.~\ref{fig:EngineeredPulses})}
\begin{tabular}{|c||c||c||}
\hline
\bf{parameter} & \bf{$\mathcal{H}^{(A)}$} & \bf{$\mathcal{H}^{(B)}$}\\[0.5ex]
\hline\hline
B-splines $N_b$                              &  10   & 10\\
carrier frequencies $N_f$                    &  22   & 17\\
$T$ guard states $\delta n_T$                &  3    & 3\\
$C$ guard states $\delta n_m$, $\delta n_l$  &  2    & 2\\
max iterations                               &  100  & 30-150\\
number of restarts                           &  10   & 10\\
target fidelity 1-$O_F$                      &  0.99 & 0.99\\
 \hline\hline
\end{tabular}
\end{table}

The optimization heuristics has a stopping condition based on either the achievement of a target threshold fidelity (1-$O_F$) or the execution of a maximum number of iterations. As mentioned, we perform multiple restarts initializing the optimization with different random pulses (see Table \ref{tab:parameters} for a summary of some of the parameters used for the numerical experiments). Computations have been performed  allowing an optimization time in the order of days. See Supplemental Material for computational details.


To give a sense of the resulting control signals that generate the QAOA circuit layers, we show the resulting Fourier transform of one engineered $d_T(t)$, $d_m(t)$, $d_l(t)$ functions in Fig.~\ref{fig:EngineeredPulses}-b, for one random seed and pulse time $\tau$ = 8000 ns, which in retrospect we know guaranteeing high fidelity of the synthesis. The angle parameters $\beta$ and $\gamma$ have been set to a fixed arbitrary value of $\pi/5$ for illustration but the qualitative features of the pulses that we are describing are preserved for different $\tau$ and angles. As evident from the plots, the scheme and parameters described above clearly generate peaks around the identified resonant frequencies corresponding to the single-boson transitions in Fig.~\ref{fig:System_and_Spectrum}. In particular, for the $d_T(t)$ controls, the highest peak corresponds to $\omega_T$ while the other equispaced peaks are centered among multiples of $\xi_{Tm}$, $\xi_{Tl}$ or integer combinations of the two energy values for the $\mathcal{H}^{(B)}$ system. For the C-mode controls, the peaking frequencies are topped by $\omega_m$, $\omega_l$ and generally $\xi_{Tm}$, $\xi_{Tl}$ plus multiples of $\xi_m$, $\xi_l$ respectively.
Clearly, each reported spectrum corresponds to a real-time microwave combination of pulses that can be crafted via an arbitrary waveform generator (AWG) in an experimental setup.

In Fig.~\ref{fig:EngineeredPulses}-c, we provide the aggregated performance of the pulse engineering approach, plotting the fidelity between the final pulse $\mathcal{U}(\tau)$ and target circuit layers (phase separation and mixing) and Hadamard gates, for different pulse times. We show the mean fidelity, estimated averaging 10 random initializations, i.e. restart of the L-BFGS optimizer (the default optimizer for \texttt{Juqbox.jl}), for QAOA layers parametrized with 11 different $\gamma$ and $\beta$ (from -$\pi$ to $\pi$ in fractions of $\pi$/5). As expected, notwithstanding outliers, the statistics is sufficient to indicate that the method can reach the target 0.99 fidelity if the pulse is allowed to be sufficiently long.

\section{Discussion and Outlook}\label{sec:discussion}

In the previous section, we described a proof-of-concept of numerical synthesis for simple quantum circuits describing the building blocks of Max-k-Cut QAOA  using qubits (mapped onto qudits) and qutrits, on bosonic quantum processors. The main question that is left to be addressed is if the synthesis approach we employed is sufficiently robust to be applied at application-scale. We break down the question in a discussion of three scalability challenges: Computational effort, realistic implementation, and circuit fidelity.

\paragraph{Computational Effort:} 
As mentioned, the computational effort required by numerical packages to obtain high-fidelity in our case study is already very significant and scales both with the Hilbert space size and with the pulse duration. This means that the proposed methodology will most certainly not be viable if straightforwardly applied to systems at large scale, although larger synthesis can be achieved if the code is optimized to leverage GPU clusters. The envisioned practical synthesis of larger circuits will necessarily need to be broken down in modules, each of which working on a subspace of the entire Hilbert space. The requirement for this modularization is that the  gate synthesized numerically in a system with few modes will have to be applied in a system with several modes and levels. The optimal gate from numerics should ideally act as an identity on the degree of freedoms that were not considered in the synthesis in order not to cause the crosstalk problem \cite{ozguler2022dynamics}. Scaling up the single-mode case $\mathcal{H}^{(A)}$  that we used will not likely be viable, since the non-local mapping onto qubits would require any gate to address the entire level structure independently from the locality of the gates, which is why we opted to synthesize the entire phase-separation circuit as opposed to the individual two-qubit gates independently. However, it is envisionable to generalize the $\mathcal{H}^{(B)}$ system adding more C-modes, i.e., considering the Hamiltonian
\begin{eqnarray}
\mathcal{H}^{(B)}_{multi}[N] &=&   
                      H_T + \sum_{j=1}^N \left[H_{m_j}
                      + H_{Tm_j}^{int}\right],                      
                      \label{eq:multi-qutrits}
\end{eqnarray} which is $\mathcal{H}^{(B)}$ for N=2. If the $\xi$ parameters of each C-mode are sufficiently separated, the peaked frequency structure of the engineered pulses suggests that it is possible that none of the peaks in the final pulses would correspond to resonances with single-boson excitations that we don't want to trigger, which would likely induce very small leakage outside the two-mode target computational space. This needs to be verified theoretically or numerically in future work. Ultimately, frequency crowding will be an issue and more sophisticated numerics or frequency spacing and bandwidth engineering will be required.

It should be noted that if the modularization works as expected, the computing time spent synthesizing algorithmic primitives would be an offline \emph{una tantum} cost to be paid to populate a lookup table (LUT) that would be accessed at runtime by the perspective user of the quantum solver. Indeed, similarly as in other domains, it is envisioned that the LUT would be computed for a large grid of parameters (angles $\gamma$ and $\beta$ in our QAOA example) and then machine learning algorithms would learn and return an interpolation of the engineered pulses if the compiler is called for a parameter that was not pre-computed, or would use nearby known points to initialize a fast optimization round to engineer a new pulse on the fly \cite{xu2022neural}.

\paragraph{Realistic Implementation:} 

While the described technique is generically applicable to any bosonic interacting system, our case study has a specific 3D cQED implementation in mind, as illustrated in the inset of Fig.~\ref{fig:System_and_Spectrum}. 
It should be noted that the general framework that we employed, pulse engineering via QOC, while proven powerful~\cite{heeres2017implementing} is not the only known approach to achieve universal synthesis of unitary quantum gates defined in the Fock space for these kind of systems. For instance, the use of selective number-dependent arbitrary phase (SNAP) protocol~\cite{heeres2015cavity,fosel2020efficient} or echoed conditional displacement~\cite{eickbusch2021fast} are strong candidates for the universal control of a single-mode system. Qudits have potential to be affected by noise less so than qubits \cite{otten2021impacts} but working with large photon-number states comes with additional complications in terms of decoherence, which are still theoretically not entirely understood~\cite{hanai2021intrinsic}.

The multiqudit system (Eq.~\ref{eq:multi-qutrits}) could be viable but its practical implementation will likely suffer from the aforementioned quantum and classical crosstalk problems whose handling is currently one of the main active research topics of the 3D multimode cQED domain~\cite{chakram2020seamless}. Even assuming that the bandwidth of the control pulses and the level spacing has sufficient resolution, there is a need for the co-design of a NISQ cQED architecture that would allow two-mode gates to operate in large Hilbert space with a controllable effect over spectator modes that are subject to an always-on interaction \cite{alam2022quantum}. Theory results on quantum adiabatic protocols~\cite{das2008colloquium, ozguler2018steering} on bosonic systems could provide an initial reference point to be generalized~\cite{pino2018quantum, starchl2022unraveling}.

\paragraph{Fidelity:} The fidelity target we used in our prototype (0.99) is in line with the fidelity of native gates in industrial grade quantum processors but it is of course somewhat arbitrary. In accordance with conservative models of uncorrelated errors, we could estimate the final fidelity of the entire circuits in Fig.~\ref{fig:EngineeredPulses}-a as the product of the fidelities of each synthesized layer, which means that ultimately the fidelity decreases exponentially with the number of layers. Hence, quantum-volumetric tests~\cite{blume2020volumetric} would fail rather fast if we were to scale our circuits beyond few variables. However, it should be noted that for quantum optimization algorithms of the variational type, it is not clear if high fidelities are required, considering that the underlying computational principle is preserved for Lindblad evolution~\cite{yang2017optimizing}. The degree of freedom of parameter setting might contribute to mitigate the misspecification of the gates due to poor synthesis. The non-requirement of exact synthesis is intuitive, since for optimization tasks we are not necessarily trying to reproduce a quantum process but rather to drive the system towards a probability distribution, which might be achievable also with partially coherent systems or in the presence of spurious unknown interactions that give rise to systematic coherent errors. So, as long as the nature of the errors is not specifically adversarial against the optimization tasks, there is still reasonable hope that a low-fidelity circuit could deliver speedup in the NISQ era. An important contribution that we are considering to improve the fidelity would be to generalize the technique of \texttt{Juqbox.jl} to open systems, and fit the experimental noise to solve for a more realistic model. Fortunately, there has already been active development in that direction, including enabling quantum optimal control and pulse-level programming in XACC \cite{nguyen2020extending, nguyen2021enabling} with \texttt{QuaC} plugin \cite{otten2017quac}, and a recently released open-source package for high-performance optimal control, \texttt{Quandary}~\cite{gunther2021quandary}.

In conclusion, we investigated the application of quantum optimal control techniques to design unitary gates for a class of physical systems that could be programmed to act as qudit-based quantum computers. We used variational algorithms such as QAOA for qubits (mapped onto a single qudit) and qutrits as targets for our case-study. Our current results, similar to other applied quantum computing works for multimode cQED~\cite{kurkcuoglu2021quantum}, are still limited on small proof-of-concept models, due to limitations in computational effort, realistic implementation and achievable fidelity. While we identified pathways to overcome such limitations, we should note that for the purpose of variational optimization there are multiple recent attempts to employ co-designed digital-analog approaches that are directly related to QOC as optimization algorithms~\cite{magann2021pulses,gokhale2019partial, choquette2021quantum}, and might not require the burdens of high-fidelity gate synthesis. We envision that our work could also contribute to those innovative methods that have already been delivering promising results.

\section*{Acknowledgments}

We thank Jens Koch, Srivatsan Chakram, Taeyoon Kim, Joshua Job, Matthew Reagor, Matthew Otten, Keshav Kapoor, Silvia Zorzetti, Sohaib Alam, Doga Kurkcuoglu and the SQMS 3D Algorithms Group and SQMS Codesign Group for discussions and feedback. We thank Adam Lyon, Jim Kowalkowski, Yuri Alexeev and Norm Tubman for their assistance on computing aspects, including support through XSEDE computational Project no. TG-MCA93S030 providing compute time at Bridges-2 of the Pittsburgh Supercomputer Center. A.B.\"O. thanks Gabriel Perdue, Adam and Jim for their guidance during his early career years. We thank Anders Petersson for his support in configuring \texttt{Juqbox.jl}. D.V. acknowledges support via NASA Academic Mission Service (NNA16BD14C). This  material  is  based  upon  work  supported  by  the U.S. Department of Energy,  Office of Science,  National Quantum  Information  Science  Research  Centers,   Superconducting  Quantum Materials  and  Systems  Center (SQMS) under contract number DE-AC02-07CH11359. We gratefully acknowledge the computing resources provided on Bebop, a high-performance computing cluster operated by the Laboratory Computing Resource Center at Argonne National Laboratory.

\bibliography{refs.bib}




\newpage

\begin{appendices}

\newpage

\setcounter{figure}{0}
\renewcommand\thefigure{S\arabic{figure}}

\newpage

\textbf{Supplementary Material}
\label{supp:SuppMat}

\newpage


\begin{figure*}[!htbp]
\begin{centering}
\includegraphics[width = 1 \textwidth]{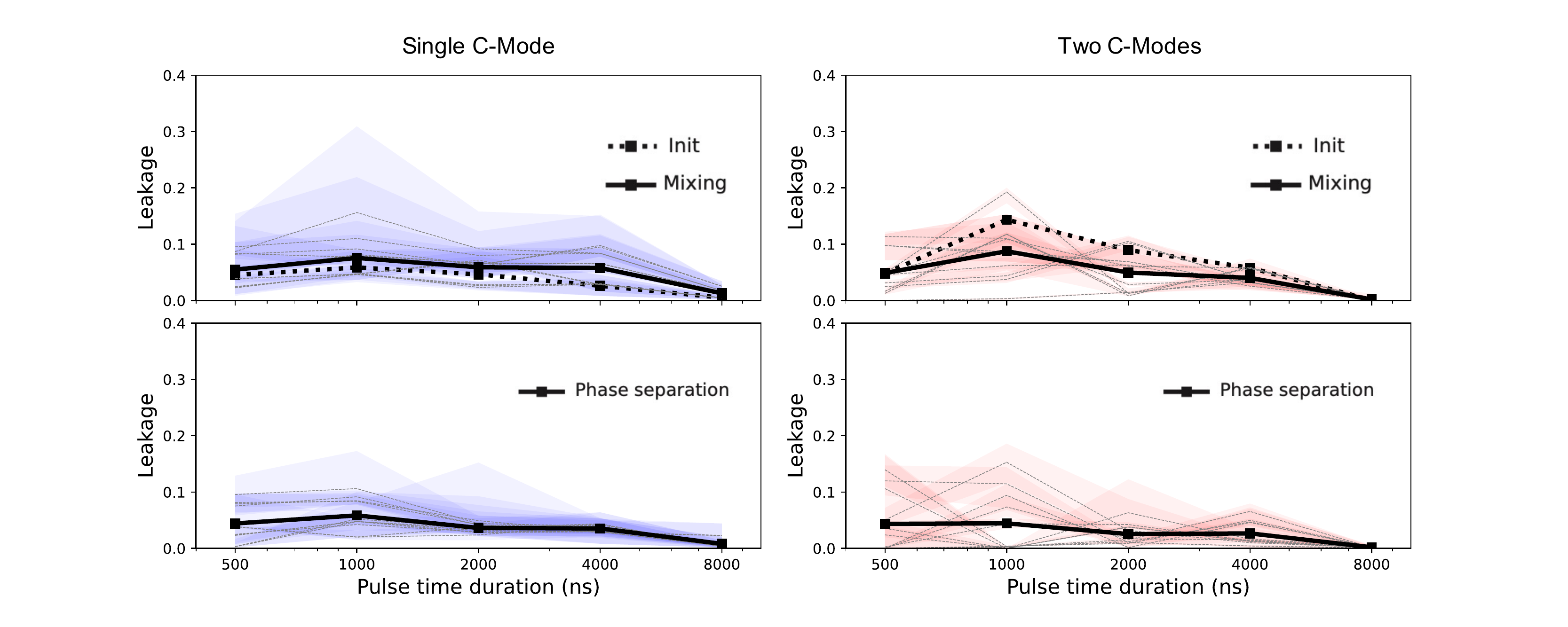}
\par\end{centering}
\caption{
Leakage (max population over all guard states) for pulse engineered QAOA layers of the prototype circuits as shown in Fig.~\ref{fig:EngineeredPulses}-a. Black lines indicate the mean across angles, individually plotted in gray. Each line is the mean of 10 random restarts (20-80 percentiles across restarts is plotted as shaded area). }
\end{figure*}

\begin{figure*}[!htbp]
\begin{centering}
\includegraphics[width = 0.7 \textwidth]{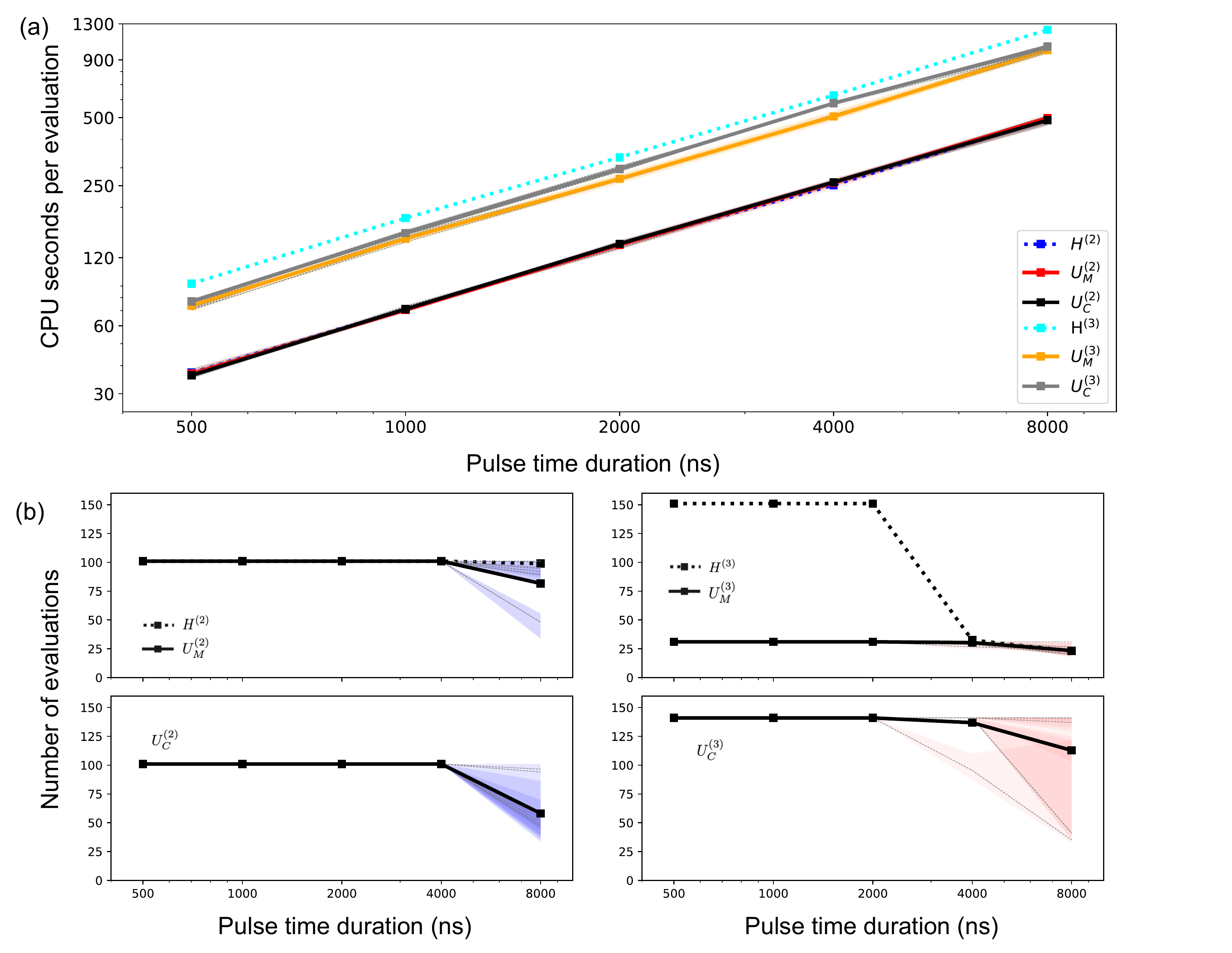}
\par\end{centering}
\caption{\label{fig:Leakage}
(a) CPU seconds per objective gradient evaluation (total CPU time divided by number of evaluations), (b) number of objective gradient evaluations. Each instance is run single-threaded on one CPU core. Black lines indicate the mean across angles, individually plotted in gray. Each line is the mean of 10 random restarts (20-80 percentiles across restarts is plotted as shaded area). Note that the shaded areas in (a) are almost negligible, which shows that standard deviation of CPU time over QAOA angles and random starts is small. (b) shows that the target fidelity is generally not reached within max number of iterations for small pulse durations. The drop at large pulse time duration happens due to reaching the target fidelity (0.99) within less than max number of iterations. Note that number of objective gradient evaluations equals (number of iterations + 1). Note also that number of max iterations are not set the same for each gate, see Table \ref{tab:parameters}.}
\end{figure*}

\end{appendices}

\end{document}